\documentclass[journal,twoside]{IEEEtran}

\usepackage[english]{babel}
\usepackage[utf8]{inputenc}
\usepackage[cmex10]{amsmath}
\usepackage{commath}
\usepackage{graphicx}
\usepackage{hyperref}
\usepackage{multirow}
\usepackage{xcolor}
\usepackage{csquotes}
\usepackage{comment}
\usepackage{lipsum}  

\ifCLASSOPTIONcompsoc
  \usepackage[caption=false,font=normalsize,labelfont=sf,textfont=sf]{subfig}
\else
  \usepackage[caption=false,font=footnotesize]{subfig}
\fi
\ifCLASSOPTIONcaptionsoff
  \usepackage[nomarkers]{endfloat}
 \let\MYoriglatexcaption\caption
 \renewcommand{\caption}[2][\relax]{\MYoriglatexcaption[#2]{#2}}
\fi

\usepackage{cite}

\newcommand{\df}{{\rm d}}

\newcommand{\nli}{\textup{NLI}}
\newcommand{\lin}{\textup{LIN}}
\newcommand{\tx}{\textup{TX}}

\begin{document}
\bstctlcite{IEEEexample:BSTcontrol}
\title{Introducing the Generalized GN-model for Nonlinear Interference Generation 
including space/frequency variations of loss/gain}
\author{M.~Cantono,~\IEEEmembership{Student~Member,~OSA},~D.~Pilori,~\IEEEmembership{Student~Member,~OSA},~A.~Ferrari,~\IEEEmembership{Student~Member,~OSA},~and~V.~Curri,~\IEEEmembership{Member,~IEEE}
\thanks{
The authors are with DET, Politecnico di Torino, 1029 Torino, Italy. E-Mail:
\{mattia.cantono,dario.pilori,alessio.ferrari,vittorio.curri\}@polito.it. 
} }
\maketitle
\begin{abstract}
We  develop and present a generalization of the GN-model --
the generalized Gaussian noise (GGN) model  --
to enabling a fair application of GN-model 
to predict generation of nonlinear interference
when loss parameters relevantly vary with frequency and/or 
distributed amplification applies selectively to portions of the exploited 
spectrum and/or stimulated-Raman-scattering-induced crosstalk is relevant.
\end{abstract}
\IEEEpeerreviewmaketitle
\section*{Paper versions}

\textbf{Version 1}
First derivation of the GGN-model with rigorous derivation starting from \cite{arXiv_GN-model}.

\textbf{Version 2}
Typos corrections including a small fix in the derivation of the GGN-reference formula. The mismatch in NLI estimation with respect to Version 1 is not relevant for reasonable power levels (up to $\sim2$ dBm/channel), i.e. where the perturbative approach is valid. 
More details with respect to the GN-model-like averaging procedures are also provided.
Updated references to other independent derivations following the same approach firstly reported in Version 1, are also reported \cite{roberts_channel_2017,semrau2018arxiv}. 
\section{Introduction}
\label{sec:intro}
\IEEEPARstart{E}{xperiments} have shown NLI generation limiting perfomances up to multi-THz spectral occupations -- as for instance in \cite{Pastorelli2012,UCL_OFC17,OE_UCL} -- 
indicating that NLI and ASE noise 
accumulation are the limiting phenomena also when exploiting the full the C-band (and possibly beyond),
at least for the experimentally tested scenarios.
Moreover, from these experimental results appears that in the analyzed scenarios the Gaussian noise (GN) model 
\cite{Carena-JLT-2012}, is an accurate yet conservative approximation of NLI generation due to Kerr effect over ultra-wide bandwidth occupation.
A wide-bandwidth use of the GN-model needs its generalization considering frequency variation of loss and gain,  including cross-talk induced by the stimulated Raman scattering (SRS).
Consequently, we introduce the Generalized GN-model (GGN) following the procedure of \cite{arXiv_GN-model}
introducing frequency/space variations of power evolution in fiber spans.

\section{Problem Definition}
GGN-model derivation follows the method already exploited in deriving the GN-model
detailed described in \cite{arXiv_GN-model}, so assuming to analyze the propagation of 
Gaussian distributed spectral components.
Then, in addition to \cite{arXiv_GN-model} hypotheses, 
we suppose frequency/space variation of power profile in fiber spans.
As in \cite{arXiv_GN-model}, we develop the model exploiting the single-polarization wave equation in fibers -- the non-linear Schrödinger equation (NLSE) -- then, we generalize the results including polarization relying on the Manakov equation (ME), i.e., on the dual-polarization NLSE with random birefringence -- and consequent PMD effect -- averaged out.
It has been shown that the ME can be reliably used far beyond its validity bandwidth in case of propagation of Gaussian distributed and depolarized signals \cite{OE_UCL}, and polarization division multiplexed signals based on multivelel modulation formats \cite{cantono_observing_2018}. This will be 
the validity scenario for GGN-model.

The non-linear Schrödinger equation (NLSE), in the frequency domain, has the following form:

\begin{equation}\label{eq:NLSE}
	\partial_z  E(z,f) = \left[ g(z,f) -j\beta(f) \right] E(z,f) + Q_{\nli}(z,f)
\end{equation}
where $z$ is the propagation direction, $E(z,f)$ is the Fourier transform of the propagating 
modal amplitude, $\partial_z$ is the partial derivative of $E(z,f)$ with respect to $z$, 
$\beta(f)$ is the dispersion coefficient, $g(z,f)$ is the profile of amplitude evolution
of modal amplitude that may vary with respect to both $z$ and $f$, $Q_{\nli}(z,f)$ 
is the non-linear term determined by the Kerr effect that is given by:
\begin{equation}\label{eq:Qnli}
	Q_{\nli}(z,f) = -j \gamma E(z,f) \ast E^\ast(z,-f) \ast E(z,f) \; .
\end{equation}
where "$\ast$" is the convolution operator.
according to the theory of differential equation, the formal solution of
Eq. (\ref{eq:NLSE}), i.e., of the evolution of the Fourier transform of the modal amplitude vs. $z$, has the following form:
\begin{equation}\label{eq:NLSEsol}
	E(z,f) = e^{\Gamma(z,f)} \int_{0}^{z} e^{-\Gamma(\zeta,f)} Q_{\nli}(\zeta,f) \df \zeta  \, + \, e^{\Gamma(z,f)}E(0,f)
\end{equation}
where $\Gamma(z,f)$ is given by:
\begin{equation}\label{eq:Gamma}
	\Gamma(z,f) = \int_{0}^{z} -j\beta(f) + g(\zeta,f) \df \zeta = -j\beta(f)z +  \int_{0}^{z} g(\zeta,f) \df \zeta
\end{equation}

We can subdivide $E(z,f)$ solution of  Eq. (\ref{eq:NLSEsol}) 
as the sum of two components: $E_{\lin}(z,f)$ considering linear propagation effects 
only ($g(z,f) -j\beta(f)$), and $E_{\nli}(z,f)$ considering the impairment of Kerr effect and its interaction with linear propagation.
Therefore, 

\begin{equation}
	E(z,f) = E_{\nli}(z,f) + E_{\lin}(z,f) \;,
\end{equation}
where the linear component $E_{\lin}(z,f)$ is:
\begin{equation}
	E_{\lin}(z,f) = e^{\Gamma(z,f)}E(0,f) \;,
\end{equation}
and the non-linear component $E_{NLI}(z,f)$ is:
\begin{equation}\label{eq:Enli}
	E_{\nli}(z,f) = e^{\Gamma(z,f)} \int_{0}^{z} e^{-\Gamma(\zeta,f)} Q_{\nli}(\zeta,f) \df \zeta
\end{equation}

In general, the  formal solution of Eq. (\ref{eq:NLSEsol}) for the 
modal amplitude evolution is useful to observe the two contributes 
-- linear and nonlinear -- to propagation impairments, but cannot be practically exploited, because the unknown function $E(z,f)$ is also in the right side term of 
Eq. (\ref{eq:NLSEsol}) being the cause of $Q_{\nli}(z,f)$ as clearly displayed by 
Eq. (\ref{eq:Qnli}).
\section{Perturbative Approach on the Non-Linear Impairment}
In $SiO_2$ fibers, strength of Kerr effect is much less intense of impairments of linear propagation, mainly given by chromatic dispersion.
So, we are legitimate to exploit a perturbative approach
for which $E_{\nli}(z,f)$ is indeed a perturbation of $E_{\lin}(z,f)$.
Consequently, we may assume that the non-linear term defined in Eq. (\ref{eq:Qnli}) 
is induced only by the linear component $E_{\lin}(z,f)$ of the 
modal amplitude $E(z,f)$.
Such an approximation yields to not considering the second-order effects,
i.e., the nonlinear effects induced by $E_{\nli}(z,f)$.
So, practically, the perturbative approach implies to use the following form for 
$Q_{\nli}(z,f)$ in place of the \emph{exact}  one of Eq. (\ref{eq:Qnli}):
\begin{equation}\label{eq:QnliPert}
	Q_{\nli}(z,f) = 
	-j \gamma E_{\lin}(z,f) \ast E_{\lin}^\ast(z,-f) \ast E_{\lin}(z,f) \; .
\end{equation}
As the convolution operator is defined as:
\begin{equation}
	x(t) \ast h(t) = \int_{-\infty}^{\infty} x(\tau) h(t-\tau) \df \tau \;,
\end{equation}
we can expand Eq. (\ref{eq:QnliPert} in the following form:
\begin{align}
	&Q_{\nli}(z,f) = \notag\\
	&=-j\gamma \left[  \int_{-\infty}^{+\infty} e^{\Gamma(z,f_1)}e^{\Gamma^{\ast}(z,f_1-f)} E(0,f_1) E^{\ast}(0,f_1-f)  \df f_1 \right]  \ast \notag\\
	& \qquad \ast \left[ e^{\Gamma(z,f)} E(0,f) \right]\\
	&= -j\gamma  \iint_{-\infty}^{+\infty} e^{\Gamma(z,f_1)}e^{\Gamma^{\ast}(z,f_1-f_2)} e^{\Gamma(z,f-f_2)} \cdot \notag\\
	& \qquad \cdot E(0,f_1) E^{\ast}(0,f_1-f_2) E(0,f-f_2)  \df f_1 \df f_2\\
	&= -j\gamma  \iint_{-\infty}^{+\infty} e^{\Gamma(z,f_1)+\Gamma^{\ast}(z,f_1-f_2)+\Gamma(z,f-f_2)} \cdot \notag\\
	& \qquad \cdot E(0,f_1) E^{\ast}(0,f_1-f_2) E(0,f-f_2)  \df f_1 \df f_2\\
	&= -j\gamma  \iint_{-\infty}^{+\infty} A(z,f) \cdot \notag\\
	& \qquad \cdot E(0,f_1) E^{\ast}(0,f_1-f_2) E(0,f-f_2)  \df f_1 \df f_2
\end{align}
where $A(z,f)$ is:
\begin{align}
	&A(z,f) = \notag \\
	&= \exp \biggl( \int_0^{z} -j [\beta(f_1)-\beta(f_1-f_2)+\beta(f-f_2)] + \notag\\
	& \qquad + [g(\zeta,f_1) + g(\zeta,f_1-f_2)+ g(\zeta,f-f_2)] \df \zeta  \biggr)
	\; .
\end{align}
Then, inserting  Eq. (\ref{eq:Gamma}) in Eq. (\ref{eq:Enli}) we get the following expression for the perturbation $E_{\nli}(z,f)$:
\begin{align}\label{eq:Enli2}
	&E_{\nli}(z,f) = \notag\\
	&= e^{-j\beta(f)z}e^{\int_0^z g(\zeta,f)\df \zeta} \cdot \notag\\
	& \qquad \cdot\int_0^z e^{j \beta(f)} e^{-\int_0^{\zeta} g(z_1,f) \df z_1} Q_{\nli}(\zeta,f) \df \zeta \notag \\
	&= e^{-j\beta(f)z}e^{\int_0^z g(\zeta,f)\df \zeta} I(z,f)
\end{align}
where $I(z,f)$ is:
\begin{align}
	&I(z,f) = \notag\\
	&= -j\gamma \iint_{-\infty}^{+\infty}E(0,f_1) E^{\ast}(0,f_1-f_2) E(0,f-f_2)  \notag\\
	&\qquad \int_0^z \exp (+j \beta(f)\zeta)  \notag\\ 
	&\qquad A(\zeta,f) \exp\left( -\int_0^{\zeta}g(z_1,f) \df z_1 \right)  \df \zeta \df f_1 \df f_2 
\end{align}
Substituting in Eq. (\ref{eq:Enli2}) the expression of $I(z,f)$ we obtain the following expression for the nonlinear perturbation introduced by the Kerr effect:
\begin{align} \label{eq:Enli_final}
	&E_{\nli}(z,f) =\notag\\
	&= e^{-j\beta(f)z}e^{\int_0^z g(\zeta,f)\df \zeta} I(z,f) \notag\\
	&\qquad -j\gamma \iint_{-\infty}^{+\infty}E(0,f_1) E^{\ast}(0,f_1-f_2) E(0,f-f_2) \notag\\
	&\qquad \int_0^z \exp (+j \beta(f)\zeta) \notag\\ 
	&\qquad A(\zeta,f) \exp\left( -\int_0^{\zeta}g(z_1,f) \df z_1 \right)  \df \zeta \df f_1 \df f_2 \; .
\end{align}
Note that Eq. (\ref{eq:Enli_final}) relies on the only approximation of Kerr effect being 
a perturbation of linear propagation and does include frequency/space variations of loss and gain as $g(\zeta,f)$ in addition to frequency variations of the 
propagation constant $\beta(f)$.
\section{NLI Power Spectral Density}
In this section, we rely on the same signal form -- depolarized and Gaussian signals -- and follow the same procedure of \cite{arXiv_GN-model} to derive the power spectral density $G_{\nli}(z,f)$ of $E_{\nli}(z,f)$, that is assumed to be a Gaussian random process. Specifically, using Eq.~16 of~\cite{arXiv_GN-model}, one can write the NLI field as 
\begin{align}
    &E_{\nli}(z,f) = -j \gamma f_0^{\frac{3}{2}} e^{-j\beta(f)z}e^{\int_0^z g(\zeta,f)\df \zeta} \cdot \notag \\
    &\sum_{i=-\infty}^{+\infty} \delta(f-if_0)\sum_{m,n,k \in \widetilde{A_i}} \sqrt{ G_{\tx}(mf_0)G_{\tx}(nf_0)G_{\tx}(kf_0)} \cdot \notag\\ 
    & \xi_m\xi_n^{\ast}\xi_k \int_0^z \exp [-\Gamma(\zeta,(m-n+k)f_0) + \Gamma(\zeta,mf_0) + \notag \\
    &+\Gamma^{\ast}(\zeta,nf_0)+\Gamma(\zeta,kf_0)]  \df \zeta
\end{align}
where $f_0$ is a divider of the symbol rate, $\xi$ is a complex Gaussian random variable and $\delta(f)$ is the Dirac delta function. $\widetilde{A}_i$ represents the set of all triples (m,n,k) such that $m-n+k=i$ and $m\neq n$ or $k\neq n$. This set identifies all non degenerate fourwave mixing components, as detailed in \cite{arXiv_GN-model}. Following the same averaging procedure of Sec. IV (D) of \cite{arXiv_GN-model}, one obtains the following expression for the single polarization expression of the power spectral density of the NLI noise, i.e.
\begin{align}
    &G_{\nli}^{sp}(z,f) = 2 \gamma^2 f_0^3 \left|\exp\left[\int_0^z g(\zeta,f)\df \zeta\right]\right|^2 \cdot \notag \\
    &\sum_{i=-\infty}^{+\infty} \delta(f-if_0)\sum_{m} \sum_{k} G_{\tx}(mf_0)G_{\tx}(kf_0) \cdot \notag \\
    & G_{\tx}((m-i+k)f_0) \int_0^z \exp [-\Gamma(\zeta,if_0) +\notag \\ 
    &+ \Gamma(\zeta,mf_0) +\Gamma^{\ast}(\zeta,(m-i+k)f_0)+\Gamma(\zeta,kf_0)]  \df \zeta
\end{align}

Similarly, considering a dual polarization signal, following the exact derivation of the previous section and the averaging procedure of Sec.IV (E) in \cite{arXiv_GN-model} one can write the PSD of the NLI noise generated by dual-polarization signals as
\begin{align}
    &G_{\nli}(z,f) = \frac{16}{27} \gamma^2 f_0^3 \left|\exp\left[\int_0^z g(\zeta,f)\df \zeta\right]\right|^2 \cdot \notag \\
    &\sum_{i=-\infty}^{+\infty} \delta(f-if_0)\sum_{m} \sum_{k} G_{\tx}(mf_0)G_{\tx}(kf_0) \cdot \notag \\
    & G_{\tx}((m-i+k)f_0) \int_0^z \exp [-\Gamma(\zeta,if_0) +\notag \\ 
    &+ \Gamma(\zeta,mf_0) +\Gamma^{\ast}(\zeta,(m-i+k)f_0)+\Gamma(\zeta,kf_0)]  \df \zeta \label{eq:discrete}
\end{align}
Then, taking the limit of Eq.~\ref{eq:discrete} for $f_0 \to 0$, such expression can be written as:
\begin{align}\label{eq:Gnli1}
	&G_{\nli}(z,f) = \notag\\
	& = \frac{16}{27} \gamma^2 \left| e^{\int_0^z g(\zeta,f)\df \zeta}  \right|^2 \cdot \notag \\
	&\qquad \cdot \iint_{-\infty}^{+\infty} G_{\tx}(f_1)G_{\tx}(f_2)G_{\tx}(f_1+f_2-f) \cdot \notag \\
	&\qquad \cdot \left| \int_0^z e^{+j [\beta(f_1+f_2-f) -\beta(f_1)+\beta(f)-\beta(f_2)]\zeta} \notag \right. \notag \cdot\\
	&\qquad \cdot \left. e^{+\int_0^{\zeta}g(z_1,f_1)-g(z_1,f)+g(z_1,f_2)+g(z_1,f_1+f_2-f) \df z_1 } \df \zeta \right|^2 \cdot \notag\\
	&\qquad \cdot \df f_1 \df f_2 \; .
\end{align}
To compact the expression, we introduce the following function $\rho(z,f)$ that considers 
the evolution of the modal amplitude vs. $z$ for each spectral component $f$:
\begin{equation} \label{eq:rho}
	\rho(z,f)=e^{\int_0^z g(\zeta,f)\df \zeta} \;.
\end{equation}
This expression may include the effect of frequency variation of loss coefficient,
of SRS-induced crosstalk and of distributed amplifications applied to
a limited portion of the exploited WDM spectrum.

Exploiting the linearity of the integral operator and the properties of the exponential function, we can rewrite the Eq. (\ref{eq:Gnli1}) as:

\begin{align}
	&G_{\nli}(z,f) = \notag\\
	&= \frac{16}{27} \gamma^2 \rho(z,f)^2 \cdot \notag \\
	&\qquad \cdot \iint_{-\infty}^{+\infty} G_{\tx}(f_1)G_{\tx}(f_2)G_{\tx}(f_1+f_2-f) \cdot \notag \\
	&\qquad \cdot \left| \int_0^z e^{+j [\beta(f_1+f_2-f) -\beta(f_1)+\beta(f)-\beta(f_2)]\zeta} \right. \cdot \notag \\
	&\qquad \cdot \left. \frac{\rho(\zeta,f_1)\rho(\zeta,f_1+f_2-f)\rho(\zeta,f_2)}{\rho(\zeta,f)}\df \zeta \right|^2 \df f_1 \df f_2 \; .
\end{align}

Finally obtaining the following expression for the NLI PSD that is the also the 
final expression of the the generalized Gaussian noise model for NLI generated by a 
single fiber span.
\begin{align} \label{eq:GGN_final}
	&G_{\nli}(z,f) = \notag\\
	&= \frac{16}{27} \gamma^2 \rho(z,f)^2 \iint_{-\infty}^{+\infty} G_{\tx}(f_1)G_{\tx}(f_2)G_{\tx}(f_1+f_2-f) \cdot \notag \\
	&\qquad \cdot \left| \int_0^z e^{+j \Delta\beta(f_1,f_2,f) \zeta} \Delta\rho(z,f,f_1,f_2) \df \zeta \right|^2 \df f_1 \df f_2
\end{align}
where $\Delta \rho$ is given by 
\begin{equation}
    \Delta\rho(z,f,f_1,f_2) =\frac{\rho(\zeta,f_1)\rho(\zeta,f_1+f_2-f)\rho(\zeta,f_2)}{\rho(\zeta,f)}
\end{equation}
The use of Eq. (\ref{eq:GGN_final}) in multi-span links is straightforward as for 
the GN-model. 
and $\Delta \beta$ 
\begin{equation}
    \Delta\beta(z,f,f_1,f_2) = [\beta(f_1+f_2-f) -\beta(f_1)+\beta(f)-\beta(f_2)]z
\end{equation}
that can be further expanded as 
\begin{equation}
    \Delta\beta(z,f,f_1,f_2) = 4\pi^2 (f_1-f)(f_2-f)[\beta_2+\pi\beta_3(f_1+f_2)]z
\end{equation}
as detailled described in Eq.~G.2 of \cite{arXiv_GN-model}. 
Eq.~\ref{eq:GGN_final} can be also expanded to be used with coherent accumulation with spans inserting 
the "phased-array" factor, or rely on the incoherent accumulation simply adding up
independently NLI generated by each fiber span. 
\section{Conclusions and future investigations}
We introduce the GGN-model in order to enable the estimate of NLI generation 
when loss/gain profiles vary with frequency.
The GGN-model can be used in case of frequency variations of loss coefficients,
frequency-varying distributed gain and SRS-induced crosstalk.
The GGN-model derivation relies on a perturbative solution of the Manakov equation
and holds also for bandwidths much larger of the ME validity bandwidth in case 
of depolarized Gaussian channels \cite{OE_UCL}, and for non Gaussian channels like shown in \cite{cantono_observing_2018}.
\bibliographystyle{IEEEtran}  
\bibliography{biblio}  
\end{document}